\documentclass[journal]{IEEEtran}
\usepackage[T1]{fontenc}
\usepackage{ifpdf}
\usepackage{cite}
\usepackage[pdftex]{graphicx}
\usepackage{amsmath}
\usepackage{amssymb}
\usepackage{bm}
\usepackage{array}
\usepackage{fixltx2e}
\usepackage{stfloats}

\usepackage[hidelinks]{hyperref}

\begin{document}
\title{Energy Efficiency Optimization: A New Trade-off Between Fairness and Total System Performance}

\author{Christos~N.~Efrem, and~Athanasios~D.~Panagopoulos,~\IEEEmembership{Senior~Member,~IEEE}
\thanks{C. N. Efrem and A. D. Panagopoulos are with the School of Electrical and Computer Engineering, National Technical University of Athens, 15780 Athens, Greece (e-mails: chefr@central.ntua.gr, thpanag@ece.ntua.gr).

This article has been accepted for publication in \textit{IEEE Wireless Communications Letters}, DOI: 10.1109/LWC.2019.2897094.  Copyright \textcopyright \ 2018 IEEE. Personal use is permitted, but republication/redistribution requires IEEE permission. See \url{http://www.ieee.org/publications_standards/publications/rights/index.html} for more information.
}}


\maketitle

\begin{abstract}
The total energy efficiency (TEE), defined as the ratio between the total data rate and the total power consumption, is considered the most meaningful performance metric in terms of energy efficiency (EE). Nevertheless, it does not depend directly on the EE of each link and its maximization leads to unfairness between the links. On the other hand, the maximization of the minimum EE (MEE), i.e., the minimum of the EEs of all links, guarantees the fairest power allocation, but it does not contain any explicit information about the total system performance. The main trend in current research is to maximize TEE and MEE separately. Unlike previous contributions, this letter presents a general multi-objective approach for EE optimization that takes into account both TEE and MEE at the same time, and thus achieves various trade-off points in the MEE-TEE plane. Due to the nonconvex form of the resulting problem, we propose a low-complexity algorithm leveraging the theory of sequential convex optimization (SCO). Last but not least, we provide a novel theoretical result for the complexity of SCO algorithms.
\end{abstract}


\IEEEpeerreviewmaketitle

\section{Introduction}

\IEEEPARstart{E}{nergy} efficiency expresses the amount of information that can be reliably transmitted per Joule of consumed energy (measured in bit/Joule), and is recently characterized as a key performance indicator for 5G networks. Zappone \textit{et al.} \cite{Zappone} propose a unified framework for the design of both centralized and distributed energy-efficient power control algorithms. Furthermore, power allocation strategies for maximizing the proportional, max-min, and harmonic fair EE in spectrum-sharing networks are given in \cite{Guo}. The optimization of various EE performance metrics is also investigated in\cite{Yang} and\cite{Venturino} for MIMO (multiple-input multiple-output) and OFDMA (orthogonal frequency division multiple access) systems, respectively. Finally, the recent study \cite{Efrem} presents a systematic approach to weighted-sum EE maximization in wireless networks.

In summary, the existing approaches maximize the\linebreak total/global, sum, product and minimum EE individually. The TEE, albeit the most important EE metric, does not depend directly on the links' EEs and its maximization results in low fairness. On the other hand, the last three EE metrics explicitly depend on the links' EEs, but none of them contains specific information about the total system performance (i.e., TEE). Moreover, the fairest resource allocation is achieved by maximizing the MEE. Consequently, in this letter, we introduce a new multi-objective approach that takes into consideration the two extremes (TEE and MEE) at the same time, and thus providing a set of MEE-TEE operating points which are not achievable with existing approaches. 

The remainder of this letter is organized as follows. Section II introduces the system model and formulates the general EE optimization problem. Subsequently, an EE optimization algorithm is developed and analyzed in Section III. Finally, numerical results are provided in Section IV, while concluding remarks are given in Section V.

\section{System Model and Problem Formulation}
We consider a wireless network with $N$ transmitters/users, $M$ receivers and $K$ mutually orthogonal resource blocks of bandwidth ${B_{RB}}$. In addition, we assume that each transmitter is associated to exactly one receiver (its intended receiver), and therefore it holds that $N \geqslant M$\footnote{Without loss of generality, we make this assumption to reduce the amount of notation needed to express the SINR in (1). Similar formula can be obtained when each transmitter is associated to more than one receiver.}. The Signal-to-Interference-plus-Noise-Ratio (SINR) experienced by user $i$ ($1 \leqslant i \leqslant N$) at its intended receiver on resource block $k$ ($1 \leqslant k \leqslant K$) is given by the following formula\footnote{The proposed methodology can be straightforwardly modified to include a self-interference term in the denominator of (1), as in \cite{Zappone} and \cite{Efrem}.}:
\begin{equation}
\gamma _i^{(k)} = {{\omega _{i,i}^{(k)}p_i^{(k)}} \mathord{\left/
 {\vphantom {{\omega _{i,i}^{(k)}p_i^{(k)}} {\left( {\sum\nolimits_{j \ne i} {\omega _{j,i}^{(k)}p_j^{(k)}}  + \mathcal{N}_i^{(k)}} \right)}}} \right.\kern-\nulldelimiterspace} {\left( {\sum\nolimits_{j \ne i} {\omega _{j,i}^{(k)}p_j^{(k)}}  + \mathcal{N}_i^{(k)}} \right)}}
\end{equation}
where $p_j^{(k)}$ is the transmit power of user $j$, $\mathcal{N}_i^{(k)}$ is the noise power at the ${i^{th}}$ user's intended receiver, and $\omega _{j,i}^{(k)}$ is the channel gain between ${j^{th}}$ transmitter and ${i^{th}}$ user's intended receiver, all on resource block $k$. For convenience, we denote the vector of transmit powers by ${\mathbf{p}} = {\left[ {{\mathbf{p}}_1^T,{\mathbf{p}}_2^T, \ldots ,{\mathbf{p}}_N^T} \right]^T}$, where ${{\mathbf{p}}_i} = {\left[ {p_i^{(1)},p_i^{(2)}, \ldots ,p_i^{(K)}} \right]^T}$ with $1 \leqslant i \leqslant N$. 

The ${i^{th}}$ user's and total achievable data rate (in bit/s) are given respectively by: ${R_i}({\mathbf{p}}) = {B_{RB}}\sum\nolimits_{k = 1}^K {{{\log }_2}\left( {1 + \gamma _i^{(k)}} \right)}$ and ${R_{tot}}({\mathbf{p}}) = \sum\nolimits_{i = 1}^N {{R_i}({\mathbf{p}})}$. Next, assuming that the transmit power amplifiers operate in the linear region and the hardware dissipated power is fixed, the ${i^{th}}$ user's and total power consumption can be modeled respectively as follows: ${P_{c,i}}({{\mathbf{p}}_i}) = {\mu _i}\sum\nolimits_{k = 1}^K {p_i^{(k)}}  + {P_{st,i}}$ and ${P_{c,tot}}({\mathbf{p}}) = \sum\nolimits_{i = 1}^N {{P_{c,i}}({{\mathbf{p}}_i})}$, where ${\mu _i} = {{1 \mathord{\left/{\vphantom {1 \xi }} \right.\kern-\nulldelimiterspace} \xi }_i}$, with ${\xi _i} \in (0,1{\kern 1pt} ]$ the efficiency of the power amplifier of transmitter $i$, and ${P_{st,i}}$ is the static dissipated power in all other circuit blocks of the ${i^{th}}$ transmitter and its intended receiver (e.g., cooling, filtering, signal up and down conversion, digital-to-analog and analog-to-digital conversion). Furthermore, the ${i^{th}}$ user's and total EE (in bit/Joule) are defined respectively as the following ratios: $E{E_i}({\mathbf{p}}) = {{{R_i}({\mathbf{p}})} \mathord{\left/{\vphantom {{{R_i}({\mathbf{p}})} {{P_{c,i}}({{\mathbf{p}}_i})}}} \right.
\kern-\nulldelimiterspace} {{P_{c,i}}({{\mathbf{p}}_i})}}$ and $E{E_{tot}}({\mathbf{p}}) = {{{R_{tot}}({\mathbf{p}})} \mathord{\left/{\vphantom {{{R_{tot}}({\mathbf{p}})} {{P_{c,tot}}({\mathbf{p}})}}} \right.\kern-\nulldelimiterspace} {{P_{c,tot}}({\mathbf{p}})}}$.

Now, we introduce the following nonconvex maximization problem, based on the multi-objective optimization theory:
\begin{equation}
\mathop {\max }\limits_{{\mathbf{p}} \in {S_{\mathbf{p}}}} \quad G({\mathbf{p}}) = F\left( {E{E_{tot}}({\mathbf{p}}),\mathop {\min }\limits_{1 \leqslant i \leqslant N} E{E_i}({\mathbf{p}})} \right)
\end{equation}
with feasible set ${S_{\mathbf{p}}} = \{ {\mathbf{p}} \in \mathbb{R}_ + ^{NK}: \; \sum\nolimits_{k = 1}^K {p_i^{(k)}}  \leqslant P_{{\kern 1pt} i}^{\max },\allowbreak \;\text{and}\;{R_i}({\mathbf{p}}) \geqslant R_i^{th}\;\text{for}\;1 \leqslant i \leqslant N\}$, where $P_{{\kern 1pt} i}^{\max }$ and $R_i^{th}$ are the ${i^{th}}$ user's maximum transmit power and minimum required data rate, respectively. Moreover, we assume that: \textit{1) the objective $F(x,y)$ is an increasing function of $x$ and $y$}, \textit{2) $F({2^u},{2^v}) > 0, \; \forall (u,v) \in {\mathbb{R}^2}$}, and \textit{3) $f(u,v) = {\log _2}F({2^u},{2^v})$ is a differentiable concave function}. 

In the sequel, we transform the original nonconvex problem (2) into an equivalent problem in a more tractable form. Due to the fact that $F(x,y)$ is an increasing function and $E{E_{tot}}({\mathbf{p}}),\mathop {\min }\limits_{1 \leqslant i \leqslant N} E{E_i}({\mathbf{p}}) \geqslant 0, \; \forall {\mathbf{p}} \in \mathbb{R}_ + ^{NK}$, problem (2) can be equivalently written as follows:
\begin{equation}
\mathop {\max }\limits_{({\mathbf{p}},\eta _{tot}^{th},\eta _{\min }^{th}) \in {\rm T} } \quad F\left( {\eta _{tot}^{th},\eta _{\min }^{th}} \right)
\end{equation}
with feasible set ${\rm T}  = \{ ({\mathbf{p}},\eta _{tot}^{th},\eta _{\min }^{th}) \in \mathbb{R}_ + ^{NK + 2}: \; {\mathbf{p}} \in {S_{\mathbf{p}}},\allowbreak \;E{E_{tot}}({\mathbf{p}}) \geqslant \eta _{tot}^{th},\;\text{and}\;E{E_i}({\mathbf{p}}) \geqslant \eta _{\min }^{th}\;\text{for}\;1 \leqslant i \leqslant N\}$, where $\eta _{tot}^{th}$ and $\eta _{\min }^{th}$ are auxiliary variables. Notice that the set of constraints $E{E_i}({\mathbf{p}}) \geqslant \eta _{\min }^{th}$ ($1 \leqslant i \leqslant N$) is equivalent to $\mathop {\min }\limits_{1 \leqslant i \leqslant N} E{E_i}({\mathbf{p}}) \geqslant \eta _{\min }^{th}$, and the maximum objective value is obtained when $E{E_{tot}}({\mathbf{p}}) = \eta _{tot}^{th}$ and $\mathop {\min }\limits_{1 \leqslant i \leqslant N} E{E_i}({\mathbf{p}}) = \eta _{\min }^{th}$. 

Subsequently, by applying the variable transformation \linebreak ${\mathbf{p}} = {2^{\mathbf{q}}}$ ($p_i^{(k)} = {2^{q_i^{(k)}}}$, $1 \leqslant i \leqslant N$ and $1 \leqslant k \leqslant K$), $\eta _{tot}^{th} = {2^u}$, $\eta _{\min }^{th} = {2^v}$, and after a few mathematical operations, we get the following nonconvex problem (note that the maximization of $F$ is equivalent to the maximization of ${\log _2}F$): 
\begin{equation}
\mathop {\max }\limits_{({\mathbf{q}},u,v) \in Z} \quad f(u,v) = {\log _2}F({2^u},{2^v})
\end{equation}
with feasible set $Z = \{ ({\mathbf{q}},u,v) \in {\mathbb{R}^{NK + 2}}: \; \sum\nolimits_{k = 1}^K {{2^{q_i^{(k)}}}}  \leqslant P_{{\kern 1pt} i}^{\max },\;{R'_i}({\mathbf{q}}) \geqslant R_i^{th},\;{\psi _i}({\mathbf{q}},v) \geqslant 0\;\text{for}\;1 \leqslant i \leqslant N, \;\text{and}\;g({\mathbf{q}},u) \geqslant 0\}$, where ${R'_i}({\mathbf{q}}) = {R_i}({2^{\mathbf{q}}})$, ${R'_{tot}}({\mathbf{q}}) = {R_{tot}}({2^{\mathbf{q}}})$, ${\psi _i}({\mathbf{q}},v) = {R'_i}({\mathbf{q}}) - {\mu _i}\sum\nolimits_{k = 1}^K {{2^{q_i^{(k)} + v}}}  - {P_{st,i}}{2^v}$, and $g({\mathbf{q}},u) = {R'_{tot}}({\mathbf{q}}) - \sum\nolimits_{i = 1}^N {{\mu _i}} \sum\nolimits_{k = 1}^K {{2^{q_i^{(k)} + u}}}  - \left( {\sum\nolimits_{i = 1}^N {{P_{st,i}}} } \right){2^u}$. 

\section{EE Optimization Algorithm}
In this section, we leverage the theory of SCO (see Appendix) so as to achieve a Karush-Kuhn-Tucker (KKT) solution for the equivalent problem (4). 

\subsection{Algorithm Design and Complexity}
In order to satisfy the properties of Theorem 1 in the Appendix, we use the following inequality with logarithms \cite{Efrem} (${\log _2}0 = - \infty$ and $0 \cdot {\log _2}0 = 0$): $A(\gamma ) = {\log _2}(1 + \gamma ) \geqslant \break a \cdot {\log _2}\gamma  + b = B(\gamma ,\gamma ')$, $\forall \gamma ,\gamma ' \geqslant 0$, where $a$, $b$ are given by:
\begin{equation}
a = {{\gamma '} \mathord{\left/
 {\vphantom {{\gamma '} {\left( {1 + \gamma '} \right)}}} \right.
 \kern-\nulldelimiterspace} {\left( {1 + \gamma '} \right)}},\;\; b = {\log _2}(1 + \gamma ') - a \cdot {\log _2}\gamma'
\end{equation}
Observe that $a \geqslant 0$, ${\left. {A(\gamma )} \right|_{\gamma  = \gamma '}} = {\left. {B(\gamma ,\gamma ')} \right|_{\gamma  = \gamma '}}$, and ${\left. {\tfrac{{dA(\gamma )}}{{d\gamma }}} \right|_{\gamma  = \gamma '}} = {\left. {\tfrac{{\partial B(\gamma ,\gamma ')}}{{\partial \gamma }}} \right|_{\gamma  = \gamma '}}$. Consequently, we can construct the following lower bounds: ${R'_i}({\mathbf{q}}) \geqslant {B_{RB}}\sum\nolimits_{k = 1}^K {\left[ {b_i^{(k)} + a_i^{(k)}{{\log }_2}\left( {\omega _{i,i}^{(k)}} \right) + a_i^{(k)}q_i^{(k)}} \right]}  - {B_{RB}}\sum\nolimits_{k = 1}^K {\left[ {a_i^{(k)}{{\log }_2}\left( {\sum\nolimits_{j \ne i} {\omega _{j,i}^{(k)}{2^{q_j^{(k)}}}} + \mathcal{N}_i^{(k)}} \right)} \right]}  = \widetilde {{R'_i}}({\mathbf{q}})$, ${R'_{tot}}({\mathbf{q}}) \geqslant \sum\nolimits_{i = 1}^N {\widetilde {{R'_i}}({\mathbf{q}})}  = \widetilde {{R'_{tot}}}({\mathbf{q}})$, ${\psi _i}({\mathbf{q}},v) \geqslant \widetilde {{R'_i}}({\mathbf{q}}) - {\mu _i}\sum\nolimits_{k = 1}^K {{2^{q_i^{(k)} + v}}}  - {P_{st,i}}{2^v} = \widetilde {{\psi _i}}({\mathbf{q}},v)$, and $g({\mathbf{q}},u) \geqslant \widetilde {{R'_{tot}}}({\mathbf{q}}) - \sum\nolimits_{i = 1}^N {{\mu _i}} \sum\nolimits_{k = 1}^K {{2^{q_i^{(k)} + u}}}  - \left( {\sum\nolimits_{i = 1}^N {{P_{st,i}}} } \right){2^u} = \widetilde g({\mathbf{q}},u)$, where $a_i^{(k)}$ and $b_i^{(k)}$ are given by (5) with $\gamma ' = {\gamma '}_i^{(k)}$. Notice that $\widetilde {{R'_i}}({\mathbf{q}})$, $\widetilde {{R'_{tot}}}({\mathbf{q}})$, $\widetilde {{\psi _i}}({\mathbf{q}},v)$, and $\widetilde g({\mathbf{q}},u)$ are all concave functions (the log-sum-exp, ${2^{x + y}}$, and ${2^x}$ are convex functions \cite{Boyd}). Based on the previous analysis, we can formulate the following convex problem which depends on the parameters $a_i^{(k)}$ and $b_i^{(k)}$:
\begin{equation}
\mathop {\max }\limits_{({\mathbf{q}},u,v) \in \Omega } \quad f(u,v) = {\log _2}F({2^u},{2^v})
\end{equation}
with feasible set $\Omega  = \{ ({\mathbf{q}},u,v) \in {\mathbb{R}^{NK + 2}}: \; \sum\nolimits_{k = 1}^K {{2^{q_i^{(k)}}}}  \leqslant P_{{\kern 1pt} i}^{\max },\;\widetilde {{R'_i}}({\mathbf{q}}) \geqslant R_i^{th},\;\widetilde {{\psi _i}}({\mathbf{q}},v) \geqslant 0\;\text{for}\;1 \leqslant i \leqslant N,\;\text{and}\;\widetilde g({\mathbf{q}},u) \geqslant 0\}$.

Algorithm 1 provides an iterative SCO procedure using the following notation: ${\boldsymbol{\sigma }} = {\left[ {{\boldsymbol{\sigma }}_1^T,{\boldsymbol{\sigma }}_2^T, \ldots ,{\boldsymbol{\sigma }}_N^T} \right]^T}$ for ${\boldsymbol{\sigma }} \in \left\{ {{\mathbf{p}},{\mathbf{q}},{\boldsymbol{\gamma }},{\boldsymbol{\gamma '}},\boldsymbol{a},\boldsymbol{b}} \right\}$, where ${{\boldsymbol{\sigma }}_i} = {\left[ {\sigma _i^{(1)},\sigma _i^{(2)}, \ldots ,\sigma _i^{(K)}} \right]^T}$ with $1 \leqslant i \leqslant N$. Based on Theorem 1 in the Appendix, Algorithm 1 monotonically increases the objective $f(u,v)$ in each iteration and, under suitable constraint qualifications, converges to a point that satisfies the KKT conditions of problem (4).

\begin{table}[!t]
\centering
\begin{tabular}{l}
\hline
\textbf{Algorithm 1.} Energy Efficiency Optimization
\\ \hline
1: Choose a tolerance ${\varepsilon} > 0$, and an initial point ${\mathbf{p}} \in {S_{\mathbf{p}}}$ \\
2: Set $l = 0$, $u = {\log _2}\left( {E{E_{tot}}({\mathbf{p}})} \right)$, $v = {\log _2}\left( {\mathop {\min }\limits_{1 \leqslant i \leqslant N} E{E_i}({\mathbf{p}})} \right)$, \\
~~~\,and ${f_0} = f(u,v)$ \\
3: \textbf{repeat} \\                                                                                                                                            
4:~~~Compute the SINR vector ${\boldsymbol{\gamma }}$ according to (1), and then the \\
~~~\space\enspace parameter vectors $\boldsymbol{a},\boldsymbol{b}$ according to (5) with ${\boldsymbol{\gamma '}} = {\boldsymbol{\gamma }}$ \\
5:~~~Solve the convex optimization problem (6) with parameters $\boldsymbol{a},\boldsymbol{b}$ \\
~~~\space\enspace in order to obtain a globally optimal solution $({{\mathbf{q}}^ * },{u^ * },{v^ * })$ \\
6:~~~Set $l = l + 1$, ${\mathbf{q}} = {{\mathbf{q}}^ * }$, $u = {u^ * }$, $v = {v^ * }$, ${\mathbf{p}} = {2^{\mathbf{q}}}$ and ${f_l} = f(u,v)$ \\
7: \textbf{until} ${{\left| {f_l} - {f_{l - 1}} \right|} \mathord{\left/{\vphantom {{\left| {f_l} - {f_{l - 1}} \right|} {\left| {f_{l - 1}} \right|}}} \right.\kern-\nulldelimiterspace} {\left| {f_{l - 1}} \right|}} < \varepsilon $ \\ \hline
\end{tabular}
\end{table}

Finally, the complexity of Algorithm 1 depends on the number of iterations until convergence as well as on the complexity of each iteration (which is mainly restricted by the optimization of a convex problem). According to Theorem 2 in the Appendix, the overall complexity of Algorithm 1 is $O\left( {\left( {{1 \mathord{\left/ {\vphantom {1 \varepsilon }} \right.\kern-\nulldelimiterspace} \varepsilon }} \right)\phi (N,K)} \right)$, where $\phi (N,K)$ is the complexity of problem (6). If this convex problem is solved by an interior-point method, then $\phi (N,K)$ is polynomial in the number of variables and constraints (which are $NK + 2$ and $3N + 1$, respectively), and thus polynomial in $N$ and $K$. Ultimately, Algorithm 1 has polynomial complexity in $N$, $K$, and $\left( {{1 \mathord{\left/{\vphantom {1 \varepsilon }} \right.\kern-\nulldelimiterspace} \varepsilon }} \right)$.

\subsection{Applications}
Afterwards, we examine two special applications of Algorithm 1, namely, the weighted product (WP) and the weighted minimum (WM) of TEE and MEE, which are respectively defined as: ${F_{WP}}(x,y) = {x^w}{y^{1 - w}}$ and ${F_{WM}}(x,y) = \min \left( {{x \mathord{\left/{\vphantom {x w}} \right.\kern-\nulldelimiterspace} w},{y \mathord{\left/{\vphantom {y {(1 - w)}}} \right.
\kern-\nulldelimiterspace} {(1 - w)}}} \right)$, with $x = E{E_{tot}}({\mathbf{p}})$ and $y = \mathop {\min }\limits_{1 \leqslant i \leqslant N} E{E_i}({\mathbf{p}})$.\linebreak Note that $w$ and $1 - w$ are the priority weights of TEE and MEE, respectively ($0 \leqslant w \leqslant 1$). Specifically, $w = 1$ corresponds to TEE maximization, while $w = 0$ corresponds to MEE maximization. Moreover, we have that: ${f_{WP}}(u,v) = w \, u + (1 - w) \, v$, and ${f_{WM}}(u,v) = \min \left( {u - {{\log }_2}w,v - {{\log }_2}(1 - w)} \right)$ since $\min ({2^r},{2^s}) = {2^{\min (r,s)}}$. Observe that ${f_{WP}}(u,v)$ and ${f_{WM}}(u,v)$ are both concave functions (the minimum of concave functions is also a concave function \cite{Boyd}). 

Concerning the WM maximization, we cannot consider the KKT conditions of problem (4) directly, since the objective ${f_{WM}}(u,v)$ is not differentiable. However, Algorithm 1 converges to a point that satisfies the KKT conditions of the following problem (equivalent epigraph form of problem (4)): $\mathop {\max }\limits_{({\mathbf{q}},u,v,t) \in \Gamma } t$ with feasible set $\Gamma  = \{ ({\mathbf{q}},u,v,t) \in {\mathbb{R}^{NK + 3}}: \; ({\mathbf{q}},u,v) \in Z, \;u - {\log _2}w \geqslant t,\;\text{and}\;v - {\log _2}(1 - w) \geqslant t\}$. This statement can be easily proved if we write problem (6) in its equivalent epigraph form: $\mathop {\max }\limits_{({\mathbf{q}},u,v,t) \in \Theta } t$ with feasible set $\Theta  = \{ ({\mathbf{q}},u,v,t) \in {\mathbb{R}^{NK + 3}}: \; ({\mathbf{q}},u,v) \in \Omega ,\;u - {\log _2}w \geqslant t, \allowbreak \;\text{and}\;v - {\log _2}(1 - w) \geqslant t\}$, and observe that the properties of Theorem 1 in the Appendix are satisfied.

\section{Numerical Results}

\begin{figure}[!t]
\centering
\includegraphics[width=3.5in]{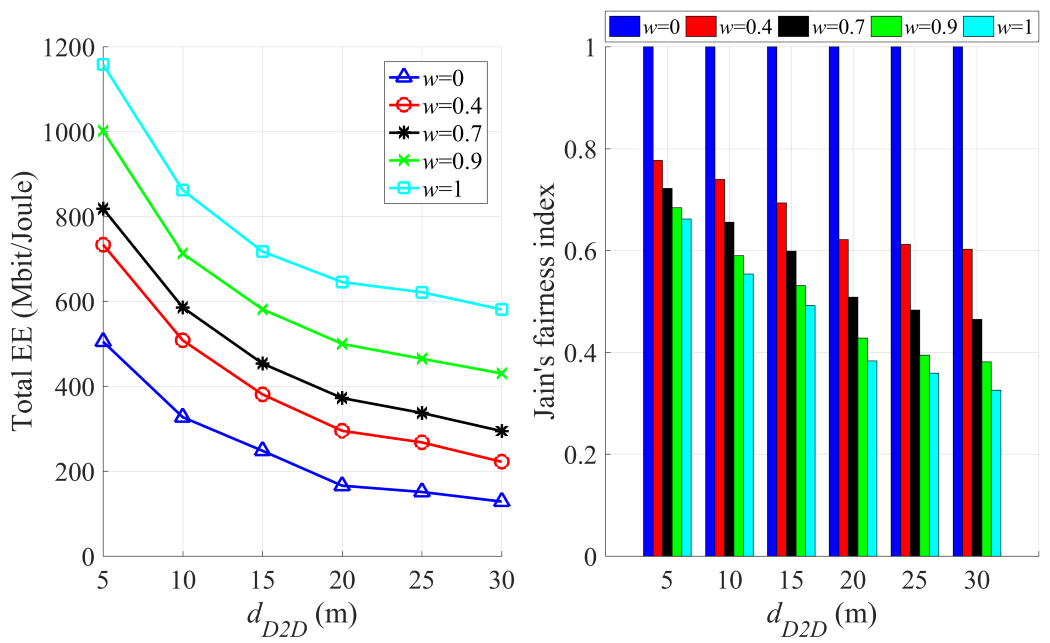} 
\caption{TEE and JFI versus D2D link distance for different priority weights.}
\label{Fig1}
\end{figure}

Consider the uplink of a cellular network with a single micro-cell, where $K = 5$ resource blocks allocated to one cellular UE (User Equipment) are reused by 4 D2D (Device-to-Device) transmitter/receiver-pairs ($N = 5$). The cellular UE is associated to the BS (Base Station) and each D2D transmitter is associated to its intended D2D receiver ($M = N$). In addition, the D2D link distance, namely, the distance between the transmitter and receiver of one D2D pair, is considered the same for all D2D pairs and is denoted by ${d_{D2D}}$. As concerns the simulation parameters, the cellular UE as well as the D2D pairs are uniformly distributed in [30,100] m from the BS. Moreover, we assume a carrier frequency of 5 GHz, $\varepsilon  = {10^{ - 3}}$, ${B_{RB}} = 500\;\text{KHz}$, $\mathcal{N}_i^{(k)} = F{\mathcal{N}_0}{B_{RB}}$ (with receiver noise figure $F = 3\;\text{dB}$, and power spectral density of the thermal noise ${\mathcal{N}_0} =  - 174\;\text{dBm/Hz}$), ${\mu _i} = \mu  = 1$, ${P_{st,i}} = {P_{st}} = 10\;\text{dBm}$, $P_{{\kern 1pt} i}^{\max } = {P_{\max }} = 23\;\text{dBm}$, and $R_i^{th} = {R_{th}} = 0$ for $1 \leqslant i \leqslant N$ (in the sequel we study the fairness, so it is preferable not to consider the data rate constraints). Unless otherwise stated, the initial feasible point is selected as ${\mathbf{p}} = \left( {{P_{\max }}/K} \right){{\mathbf{1}}_{NK \times 1}}$, where ${{\mathbf{1}}_{NK \times 1}}$ is the $NK \times 1$ vector of ones. Furthermore, all the results (except for Fig. 2) are obtained by averaging over ${10^3}$ independent simulations, and the following analysis refers to Algorithm 1 specialized to maximize the WP of TEE and MEE.

For the evaluation of fairness, we make use of Jain's fairness index (JFI) as a function of users' EEs: $\mathcal{J} = \tfrac{{{{\left( {\sum\nolimits_{i = 1}^N {E{E_i}} } \right)}^2}}}{{N\sum\nolimits_{i = 1}^N {EE_i^2} }}$ with $0 \leqslant \mathcal{J} \leqslant 1$. In general, the closer JFI is to 1, the fairer the power allocation is in terms of EE. In the special case where $w = 0$ (MEE maximization) all the EEs are equal at the maximum point \cite{Bjornson}, and therefore $\mathcal{J} = 1$ and TEE=MEE. 

First of all, Fig. 1 shows the TEE and JFI versus the D2D link distance for different weights. For fixed ${d_{D2D}}$, it is clear that TEE increases while JFI decreases as the weight $w$ increases, since higher priority is given to TEE and lower to MEE. According to the left figure, TEE decreases with the D2D link distance for all $w$. In addition, as shown in the right figure, JFI decreases with the D2D link distance for $w \ne 0$, whereas it remains equal to 1 for $w = 0$ as already mentioned.

\begin{figure}[!t]
\centering
\includegraphics[width=3.43in]{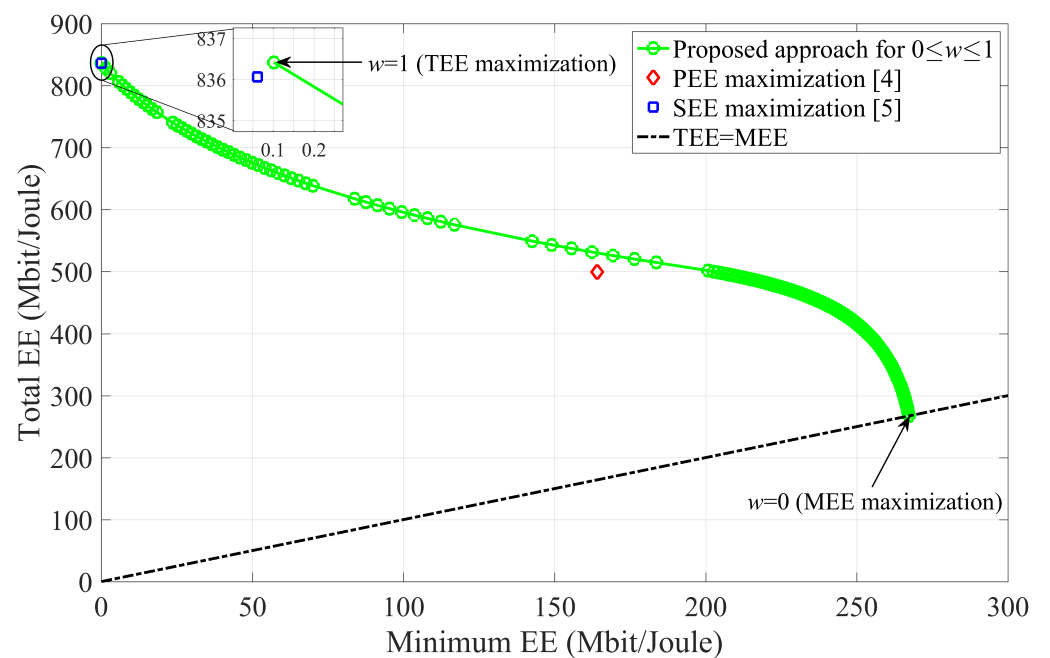} 
\caption[caption]{Pareto operating points in the MEE-TEE plane for a specific simulation scenario with ${d_{D2D}} = 10\;\text{m}$.}
\label{Fig2}
\end{figure}

\begin{figure}[!t]
\centering
\includegraphics[width=3.4in]{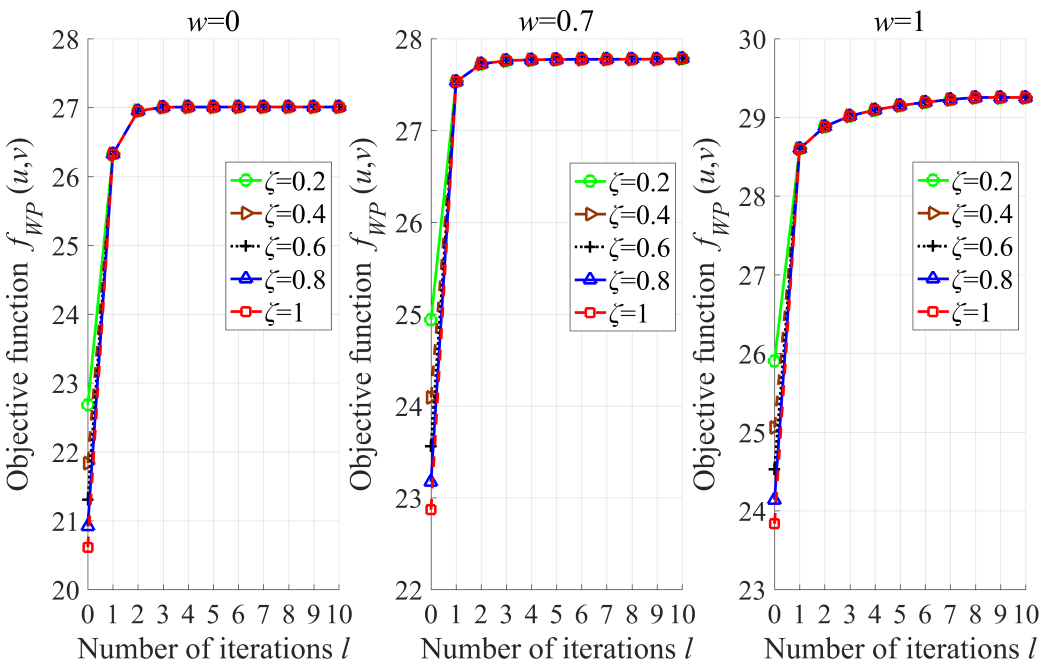} 
\caption{Convergence of Algorithm 1 (WP maximization) for different priority weights and initial point ${\mathbf{p}} = \zeta \left( {{P_{\max }}/K} \right){{\mathbf{1}}_{NK \times 1}}$ with ${d_{D2D}} = 20\;\text{m}$.}
\label{Fig3}
\end{figure}

Afterwards, Fig. 2 illustrates the Pareto operating points in the MEE-TEE plane achieved by: a) the proposed approach for 200 equally-spaced values of the weight $w$ in $[0,1]$, b) product-EE maximization\cite{Venturino} with $PEE({\mathbf{p}}) = \prod\nolimits_{i = 1}^N {E{E_i}({\mathbf{p}})}$, and c) sum-EE maximization\cite{Efrem} with $SEE({\mathbf{p}}) = \sum\nolimits_{i = 1}^N {E{E_i}({\mathbf{p}})}$. As can be seen, the proposed approach for $0<w<1$ achieves several trade-off points which are not attainable by maximizing TEE, SEE, PEE and MEE individually. Moreover, we can observe that all the Pareto points lie on or above the line TEE=MEE, since it can be easily proved that \linebreak $E{E_{tot}}({\mathbf{p}}) \geq  \mathop {\min }\limits_{1 \leqslant i \leqslant N} E{E_i}({\mathbf{p}})$, $\forall {\mathbf{p}} \in \mathbb{R}_ + ^{NK}$.

Finally, we examine the convergence of Algorithm 1 for different priority weights and initial points. According to Fig. 3, Algorithm 1 exhibits fast convergence and insensitivity to initial points for all simulation scenarios, and requires a quite small number of iterations to converge. In particular, given the tolerance $\varepsilon  = {10^{-3}}$ ($\varepsilon  = {10^{-4}}$), it converges within approximately 4, 5 and 9 (5, 6 and 10) iterations for $w = 0,\ 0.7\ \text{and}\ 1$, respectively.

\section{Conclusion}
In this letter, we have developed a unified methodology for EE optimization that incorporates a new trade-off between fairness and total system performance. Furthermore, an efficient SCO algorithm has been proposed which can be applied to practical scenarios of wireless networks. Finally, we have presented a general complexity analysis for SCO algorithms.

\appendices
\section*{Appendix \linebreak Sequential Convex Optimization}
Let $\mathcal{F}$ be a nonconvex maximization problem with objective ${f_0}({\mathbf{x}})$, and nonempty, compact feasible set $S = \{ {\mathbf{x}} \in {\mathbb{R}^n}:\;{f_i}({\mathbf{x}}) \geqslant 0,\; 1 \leqslant i \leqslant I\}$. Also, let ${\left\{ {{\mathcal{H}_j}} \right\}_{j \geqslant 1}}$ be a sequence of convex maximization problems with objective ${h_{0,j}}({\mathbf{x}},{\mathbf{x}}_{j - 1}^ * )$, compact feasible set ${S_j} = \{ {\mathbf{x}} \in {\mathbb{R}^n}:\;{h_{i,j}}({\mathbf{x}},{\mathbf{x}}_{j - 1}^ * ) \geqslant 0,\; 1 \leqslant i \leqslant I\}$, and global maximum ${\mathbf{x}}_j^ *$. Let ${\mathbf{x}}_0^ *$ be any feasible point of problem $\mathcal{F}$, that is, ${\mathbf{x}}_0^ * \in S$. Moreover, assume that ${f_i}({\mathbf{x}})$ and ${h_{i,j}}({\mathbf{x}},{\mathbf{x}}_{j - 1}^ *)$, $0 \leqslant i \leqslant I$ and $j \geqslant 1$, are differentiable functions. The next theorem follows directly from \cite{Marks}.

\vspace{2mm}
\noindent
\textbf{Theorem 1} (Convergence)\textbf{.} \textit{Suppose that the functions ${h_{i,j}}({\mathbf{x}},{\mathbf{x}}_{j - 1}^ * )$, $0 \leqslant i \leqslant I$ and $j \geqslant 1$, satisfy the following three properties (note that $\nabla  = {[ {\partial / \partial {x_1},\partial / \partial {x_2}, \ldots ,\partial / \partial {x_n}} ]^T}$): \\ 
$(a)$ ${h_{i,j}}({\mathbf{x}},{\mathbf{x}}_{j - 1}^ * ) \leqslant {f_i}({\mathbf{x}}),\; \forall {\mathbf{x}} \in {S_j}$ \\
$(b)$ ${\left. {{h_{i,j}}({\mathbf{x}},{\mathbf{x}}_{j - 1}^ * )} \right|_{{\mathbf{x}} = {\mathbf{x}}_{j - 1}^ * }} = {f_i}({\mathbf{x}}_{j - 1}^ * )$ \\
$(c)$ ${\left. {\nabla {h_{i,j}}({\mathbf{x}},{\mathbf{x}}_{j - 1}^ * )} \right|_{{\mathbf{x}} = {\mathbf{x}}_{j - 1}^ * }} = \nabla {f_i}({\mathbf{x}}_{j - 1}^ * )$ \\
Then, the sequence ${\left\{ {{f_0}({\mathbf{x}}_j^ * )} \right\}_{j \geqslant 0}}$ is monotonically increasing (${f_0}({\mathbf{x}}_j^ * ) \geqslant {f_0}({\mathbf{x}}_{j - 1}^ * ), \; j \geqslant 1$) and converges to a finite value $L$ ($\mathop {\lim }\limits_{j \to \infty } {f_0}({\mathbf{x}}_j^ * ) = L  < \infty $). In addition, every accumulation/limit point ${\mathbf{\bar x}}$ of the sequence  ${\left\{ {{\mathbf{x}}_j^ * } \right\}_{j \geqslant 0}}$  achieves the objective value $L$ (${f_0}({\mathbf{\bar x}}) = L$) and, assuming suitable constraint qualifications, satisfies the KKT conditions of the initial problem $\mathcal{F}$.}
\vspace{2mm}

A rigorous mathematical analysis for the complexity of SCO is very challenging since the convergence rate depends on the particular structure of the problem, and no theoretical results are available so far. Nevertheless, we provide the following general theorem exploiting the monotonicity of SCO.

\vspace{2mm}
\noindent
\textbf{Theorem 2} (Complexity)\textbf{.} \textit{Assume that: 1) the properties of Theorem 1 are satisfied, 2) SCO terminates when ${{\left| {{f_0}({\mathbf{x}}_j^ * ) - {f_0}({\mathbf{x}}_{j - 1}^ * )} \right|} \mathord{\left/
 {\vphantom {{\left| {{f_0}({\mathbf{x}}_j^ * ) - {f_0}({\mathbf{x}}_{j - 1}^ * )} \right|} {\left| {{f_0}({\mathbf{x}}_{j - 1}^ * )} \right|}}} \right. \kern-\nulldelimiterspace} {\left| {{f_0}({\mathbf{x}}_{j - 1}^ * )} \right|}} < \varepsilon $, where $\varepsilon  > 0$ is a predefined tolerance, and 3) ${f_0}({\mathbf{x}}_0^ * ) > 0$. Then, the number of iterations until convergence is $O\left( {{1 \mathord{\left/{\vphantom {1 \varepsilon }} \right.\kern-\nulldelimiterspace} \varepsilon }} \right)$, and  the overall complexity of SCO is $O\left( {\left( {{1 \mathord{\left/{\vphantom {1 \varepsilon }} \right.\kern-\nulldelimiterspace} \varepsilon }} \right)\varphi (n,I)} \right)$, where $\varphi (n,I)$ is the complexity of the method used to solve each convex problem with $n$ variables and $I$ constraints.}
\vspace{2mm}

\textit{Proof:} By virtue of Theorem 1, we have that ${f_0}({\mathbf{x}}_j^ * ) \geqslant {f_0}({\mathbf{x}}_{j - 1}^ * ) \geqslant {f_0}({\mathbf{x}}_0^ * ) > 0, \; j \geqslant 1$. Next, let $k \geqslant 1$ be the number of iterations until convergence, that is, the smallest integer for which ${{\left( {{f_0}({\mathbf{x}}_k^ * ) - {f_0}({\mathbf{x}}_{k - 1}^ * )} \right)} \mathord{\left/{\vphantom {{\left( {{f_0}({\mathbf{x}}_k^ * ) - {f_0}({\mathbf{x}}_{k - 1}^ * )} \right)} {{f_0}({\mathbf{x}}_{k - 1}^ * )}}} \right.\kern-\nulldelimiterspace} {{f_0}({\mathbf{x}}_{k - 1}^ * )}} < \varepsilon $. Hence, before the termination of the algorithm, it holds that $\varepsilon  \leqslant {{\left( {{f_0}({\mathbf{x}}_j^ * ) - {f_0}({\mathbf{x}}_{j - 1}^ * )} \right)} \mathord{\left/{\vphantom {{\left( {{f_0}({\mathbf{x}}_j^ * ) - {f_0}({\mathbf{x}}_{j - 1}^ * )} \right)} {{f_0}({\mathbf{x}}_{j - 1}^ * )}}} \right.\kern-\nulldelimiterspace} {{f_0}({\mathbf{x}}_{j - 1}^ * )}} \leqslant {{\left( {{f_0}({\mathbf{x}}_j^ * ) - {f_0}({\mathbf{x}}_{j - 1}^ * )} \right)} \mathord{\left/{\vphantom {{\left( {{f_0}({\mathbf{x}}_j^ * ) - {f_0}({\mathbf{x}}_{j - 1}^ * )} \right)} {{f_0}({\mathbf{x}}_0^ * )}}} \right. \kern-\nulldelimiterspace} {{f_0}({\mathbf{x}}_0^ * )}}$, and thus $\varepsilon {f_0}({\mathbf{x}}_0^ * ) \leqslant {f_0}({\mathbf{x}}_j^ * ) - {f_0}({\mathbf{x}}_{j - 1}^ * )$ for $1 \leqslant j \leqslant k - 1$ (if $k = 1$, there is no such $j$). Now, by taking the sum from $j = 1$ to $k - 1$, we get $\sum\nolimits_{j = 1}^{k - 1} {\varepsilon {f_0}({\mathbf{x}}_0^ * )}  \leqslant \sum\nolimits_{j = 1}^{k - 1} {{f_0}({\mathbf{x}}_j^ * )}  - \sum\nolimits_{j = 1}^{k - 1} {{f_0}({\mathbf{x}}_{j - 1}^ * )}$ $ \Rightarrow $ $(k - 1)\varepsilon {f_0}({\mathbf{x}}_0^ * ) \leqslant \sum\nolimits_{j = 1}^{k - 1} {{f_0}({\mathbf{x}}_j^ * )}  - \sum\nolimits_{j = 0}^{k - 2} {{f_0}({\mathbf{x}}_j^ * )}  = {f_0}({\mathbf{x}}_{k - 1}^ * ) - {f_0}({\mathbf{x}}_0^ * )$. Due to Property $(a)$ of Theorem 1, every feasible point of problem ${\mathcal{H}_j}$ is also feasible for problem $\mathcal{F}$ (${S_j} \subseteq S, \; j \geqslant 1$), and therefore ${f_0}({\mathbf{x}}_j^ * ) \leqslant {f_0}({{\mathbf{x}}^ * })$ for $j \geqslant 0$ (${{\mathbf{x}}^ *}$ is a global maximum of problem $\mathcal{F}$). This implies that ${f_0}({\mathbf{x}}_{k - 1}^ * ) \leqslant {f_0}({{\mathbf{x}}^ * })$, and thus $(k - 1)\varepsilon {f_0}({\mathbf{x}}_0^ * ) \leqslant {f_0}({{\mathbf{x}}^ * }) - {f_0}({\mathbf{x}}_0^ * )$ $\Rightarrow$ $k \leqslant 1 + {{(\lambda  - 1)} \mathord{\left/{\vphantom {{(\lambda  - 1)} \varepsilon }} \right.\kern-\nulldelimiterspace} \varepsilon } = O\left( {{1 \mathord{\left/{\vphantom {1 \varepsilon }} \right.\kern-\nulldelimiterspace} \varepsilon }} \right)$, where $\lambda  = {{{f_0}({{\mathbf{x}}^ * })} \mathord{\left/{\vphantom {{{f_0}({{\mathbf{x}}^ * })} {{f_0}({\mathbf{x}}_0^ * )}}} \right.\kern-\nulldelimiterspace} {{f_0}({\mathbf{x}}_0^ * )}} \geqslant 1$. Since the number of iterations until convergence is $O\left( {{1 \mathord{\left/ {\vphantom {1 \varepsilon }} \right. \kern-\nulldelimiterspace} \varepsilon }} \right)$, and in each iteration a convex problem is solved with complexity $\varphi (n,I)$, the thesis follows immediately.  \hfill  $\blacksquare$

In general, a global optimum of a convex problem can be obtained in polynomial time, using standard convex optimization techniques such as interior-point methods \cite{Boyd} (i.e., $\varphi (n,I)$ is a polynomial function of $n$ and $I$). Note that the best upper-complexity-bound for a generic convex problem, known so far, is $O(n^4)$ and is yielded by interior-point methods\cite{Ben-Tal}. Hence, \textit{the overall complexity of SCO is polynomial in $n$, $I$, and $\left( {{1 \mathord{\left/ {\vphantom {1 \varepsilon }} \right.\kern-\nulldelimiterspace} \varepsilon }} \right)$}.


\end{document}